\documentstyle[aps,epsfig,multicol,subfigure,color]{revtex}
\newcommand{\bcols}{\ifpreprintsty\else\begin{multicols}{2}\fi}
\newcommand{\ecols}{\ifpreprintsty\else\end{multicols}\fi}

\begin{document}
\draft

\title {Electron transport in nanotube--molecular wire hybrids} 
\author {G. Fagas, G. Cuniberti, and K. Richter}
\address {Max-Planck-Institut f{\"u}r Physik komplexer Systeme,  N{\"o}thnitzer Strasse 38, D-01187 Dresden}
\date{\today}
\maketitle
\begin{abstract}
We study contact effects on electron transport across a molecular wire sandwiched between two
semi-infinite~(carbon) nanotube leads as a model for nanoelectrodes. Employing
the Landauer scattering matrix approach we find that the
conductance is very sensitive to parameters such as the coupling strength and geometry of the contact. 
The conductance exhibits markedly different behavior
in the two limiting scenarios of single contact and multiple contacts
between the molecular wire and the nanotube interfacial atoms.
In contrast to  a single contact the multiple-contact configuration acts as a filter selecting
single transport channels. It exhibits a scaling law for the conductance as a function of 
coupling strength and tube diameter. We also observe an unusual
narrow-to-broad-to-narrow behavior of conductance resonances upon decreasing the coupling.
\end{abstract}
\bcols

The `top-down' miniaturization of electronic devices to micro- or nanometer-size has triggered
formidable progress in the research of mesoscopic systems for more than a decade now.
Such endeavor has given rise to a variety of nontrivial quantum effects in
solid state physics at these length scales providing an
active field of interplay between fundamental and applied research.

An arena closely related to mesoscopics is molecular electronics\cite{AR98}
which combines molecular and solid state physics. Driven by possible technological 
applications and owing to recent experimental breakthrough, this area has evolved 
to a rapidly developing field on its own posing new challenges to theory.
Molecular electronics is based on the `bottom-up' manufacture philosophy whose
underlying principle is to use molecules or supramolecular structures
as (reproducible) circuit elements. 
Hence, employing the diverse electronic properties of molecular complexes
and the capacity of synthetic (bio)chemistry, it may be able to circumvent inherent 
limitations and imperfections of conventional semiconductor device fabrication techniques
at nanometer scales.

Although the original idea is quite old, significant progress
has only been demonstrated experimentally in recent years. 
Owing to the advances
in self-assembly techniques \cite{ADWBCTRYKJ98}, end-group modifications \cite{Tour96},
scanning probe \cite{BR99} and break-junction techniques \cite{RZMBT97}, atomic-scale control and positioning of single
molecules and their assemblies become possible. 
First electron transport measurements
through molecular complexes between metallic electrodes have been reported.
Proposals and studies of molecular wires range from `simple' 
molecules~\cite{RZMBT97} to DNA strands~\cite{FS99}.
In a parallel development the use
of carbon nanotube networks has been the focus of intense experimental and theoretical activity
as another promising direction for building blocks of molecular circuits
\cite{RKJTCL00}.

Albeit a molecular device is typically divided into three parts, the donor and acceptor electrodes
and the molecular compound serving as a bridge, it is clear that to understand conductance
measurements an account of the system as a whole is required. This is
intuitively plausible when looking at the problem as an electron transfer process. The coupling of the molecular complex to the environment as well as its intrinsic ability to convey charge
are equally important factors. Hence, whereas the molecular character has been
the main
focus~\cite{MKR94a,STDHK96,JV96},
the precise nature of the contact and its implications
has also become a topic of investigation~\cite{EK98a,YR98}.

Yet the electrodes are usually formed from bulk material. 
In contrast, we take here the viewpoint that the electrodes in the vicinity of  the molecular interface can be mesoscopic themselves.
To be more specific, we focus on carbon tubules as suitable candidates for such nanoelectrodes. 
Carbon nanotubes are known to exhibit a wealth of properties depending on their diameter~($\sim nm$), chirality~(orientation of graphene roll up), and whether they consist of a single cylindrical surface~(single-wall) or more~(multi-wall)
~\cite{SDD98,McEuen00}. 
On the one hand, first experimental attempts to build
nanotube-molecule-nanotube hybrids are on their way~\cite{bachtoldprivate}. 
On the other hand, carbon nanotubes are utilized as scanning probe tips to study molecular structures~\cite{WJWCL98}. 
This represents a related setup where contact effects of a
molecule-nanoelectrode junction play a key role.

In the present study, we address the influence of the molecular wire-electrode contact on
the conductance for the class of systems
where the structure of electrodes plays an important role.
For mesoscopic leads with reduced dimensionality
lateral to the current direction, it makes sense to
discriminate between electron transport channels, e.g. carbon nanotubes support
up to two channels for electrons with energy around the equilibrium Fermi energy. 
Evidently, for such low-dimensional transport the geometry of the contact should crucially determine the measured conductance. We find that electron transport shares distinct properties
depending on the number and strength of contacts between the molecular bridge and the
interface as well as on the symmetry of the channel wavefunctions transverse to transport.
We demonstrate that single contacts give rise to complex conductance
spectra exhibiting quantum features of both the molecule and the electrodes;
multiple contacts provide a mechanism for transport channel selection, leading
to a scaling law for the conductance and allowing for its control.
Channel selection also highlights the role of molecular resonant states
by suppressing details assigned to the electrodes.
Such information may be used as a guideline for systematic chemical synthesis or complementary
experimental analysis.

We shall now specify more the system we have in mind. The
electrodes are open-ended single-wall
carbon nanotubes described by a parameterized tight-binding Hamiltonian with
a $\pi$-electron per
atom which represents a good approximation to the carbon nanotubes electronic band
structure~\cite{SDD98}, whereas the molecular system sandwiched in between is
modeled by a homogeneous tight-binding chain.
The latter choice reflects our aim to provide a qualitative understanding 
of the phenomenology of the physical problem as opposed to more quantitative
quantum-chemical methods that take into account the precise structure and properties
of the molecular bridge. 
\vspace*{-7mm} 
\begin{figure}[t]
\begin{center}
\subfigure{\epsfig{file=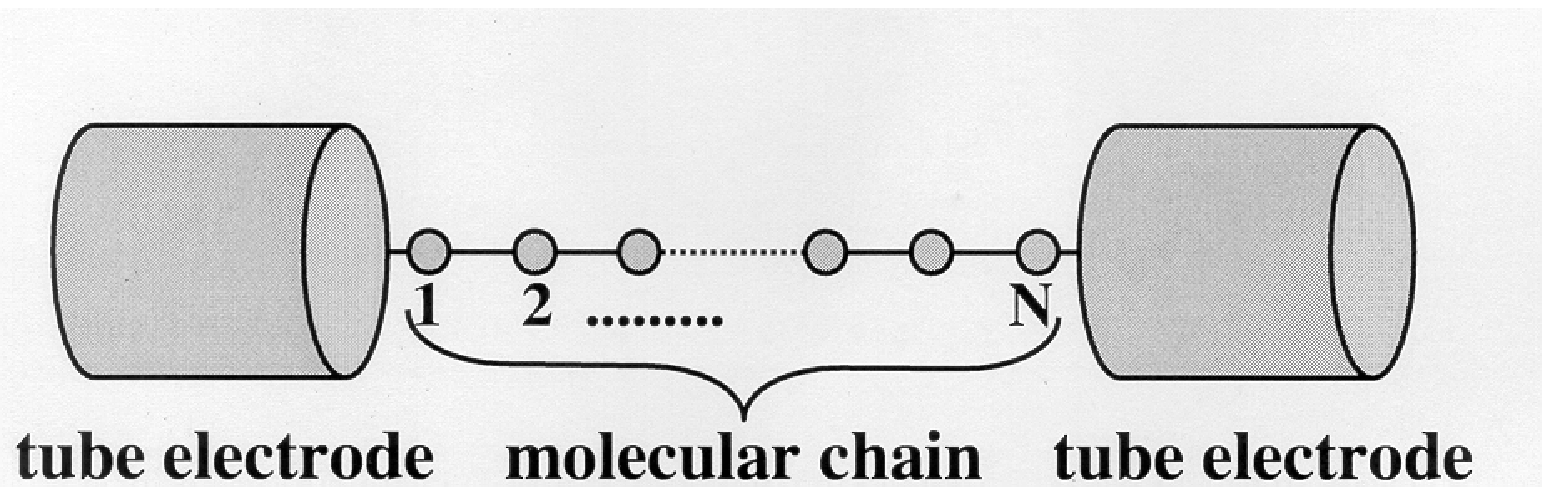, width=.49\linewidth}}
\subfigure{\epsfig{file=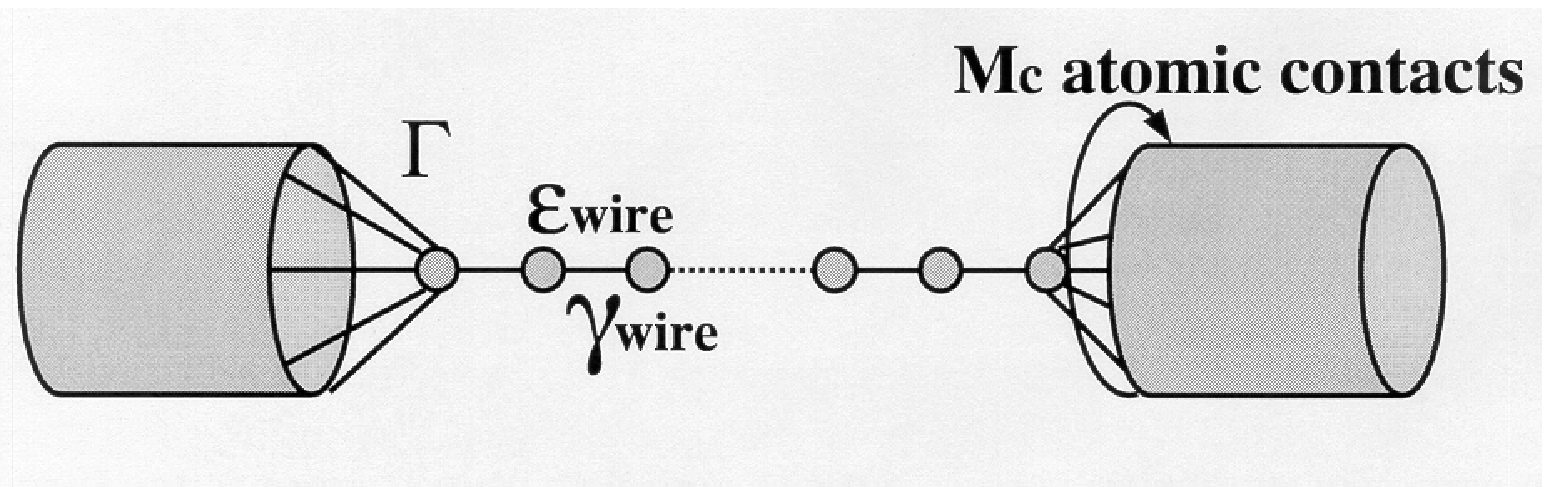, width=.49\linewidth}}
\end{center}
\vspace*{-5mm} 

\caption{\label{fig0}
	Scheme of the molecular wire-tube hybrid with
	single (left) and multiple (right) contacts.
	}
\end{figure}
The electronic Hamiltonian of the full system, including the left~(L) and
right~(R) tube (see Fig.~\ref{fig0}), reads
\begin{eqnarray}
\label{meq1}
 H &=& H_{\rm tubes} + H_{\rm wire} + H_{\rm coupling}
\\ \nonumber &=&
\sum_{\alpha={\rm L,R,wire}} \; \sum_{n^{\phantom{\prime}}_{\alpha},n_{\alpha}^\prime} 
\left( \frac{\varepsilon^{\alpha}_{n_{{\alpha}}}} 2 \delta_{n^{\phantom{\prime}}_{{\alpha}},n_{{\alpha}}^\prime}
+
 \gamma^\alpha_{\left\langle n^{\phantom{\prime}}_{\alpha},n_{\alpha}^\prime \right\rangle}
 \right)
\left| n^{\phantom{\prime}}_{\alpha} \right\rangle 
\left\langle n_{\alpha}^\prime \right| 
\\ & + & \sum_{m_{\rm L}} \Gamma \left| m_{\rm L}\right. \left.\!\! \rangle 
 \langle n_{\rm wire}\!=\! 1\right.\left.\!\! \right|
 \nonumber
 + \sum_{m_{\rm R}} \Gamma \left| m_{\rm R}\right. \left. \! \! \rangle 
 \langle n_{\rm wire}\!=\! N\right.\left.\!\! \right|
+ {\rm H.c.} . \nonumber
\end{eqnarray}
Here, $\gamma^{\rm L,R}$($=2.66$ eV), $\gamma^{\rm wire}$, and $\Gamma$ are the 
hopping matrix elements between atoms of the carbon tubule leads, molecular bridge, and
the bridge/lead interface, respectively. 
They are non-zero only for nearest neighbors.
In Eq.~(\ref{meq1}), $\varepsilon^{\rm wire}$ is the on-site or orbital energy of each of the $n_{\rm wire}=1,\dots, N$
chain-atoms relative to that of the leads, $\varepsilon^{\rm L,R}$, which is fixed to zero.
Summations over $m_{\rm L}$ and $m_{\rm R}$ run over interfacial end-atoms of the leads.
In general, there are $M$ such atomic positions, depending on the perimeter
of the tubes, and $1\leq M_{\rm c} \leq M$ hybridization contacts.
We also compare with a square lattice model of 
mesoscopic electrodes with nearest-neighbor 
interactions~($\gamma^{\rm L,R}\!=\!1$eV) and periodic boun\-dary conditions,
which delivers additional insight.

In what follows, we use the Landauer theory \cite{IL99}
which relates the conductance
of a system
to an
independent-electron scattering problem~\cite{f1} and describes
unique quantum effects
in mesoscopic systems~\cite{FG99}.
The electron wavefunction is assumed to extend coherently across the device
and the two-terminal, linear-response conductance at zero temperature reads
\begin{equation}
\label{meq2}
G(E_{\rm F}) = 2 (e^2/h) T(E_{\rm F}).
\end{equation}
The factor two accounts for spin degeneracy, and $T(E_{\rm F})$ is the total
transmittance 
for injected electrons with Fermi energy $E_{\rm F}$.
The transmission function is given by
$T(E)=\sum_{\nu\nu^\prime}|S_{\nu\nu^\prime}(E)|^2$, where $\nu,\nu^\prime$ are
quantum numbers labeling open channels for transport which belong to mutually
exclusive leads, in our case the two semi-infinite perfect nanotubes.
The molecular system attached acts as a scatterer, and $S$ is the 
corresponding quantum-mechanical scattering matrix.

For the numerical calculation of the central quantity, $T(E)$,
we use a general scattering technique which has been recently formulated for
studies of the giant magnetoresistance~\cite{SKTL00}. Application of the method
to phonon transport across disordered interfaces~\cite{FKLWPPdenH99} and to
electric conductance in multi-wall
carbon nanotubes~\cite{SKTL00} reveals an efficient algorithm
for calculating the Green function for arbitrary tight-binding Hamiltonians
and, hence, the $S$-matrix~\cite{FG99}.
The computational scheme comprises two essential steps:
first, the calculation of an effective~(renormalized) interaction between the
electrodes by projecting out the degrees of freedom of the scatterer,
and second, the computation of the unperturbed
electrodes Green function. Then, one uses the Dyson equation to express the
Green function of the composite system (leads plus scatterer). 
We followed the suggestion discussed in \cite{SKTL00} for
the implementation and computation 
of the exact Green function of the tubules.
\vspace*{-5mm} 
\begin{figure}
\begin{center}
\subfigure{\epsfig{file=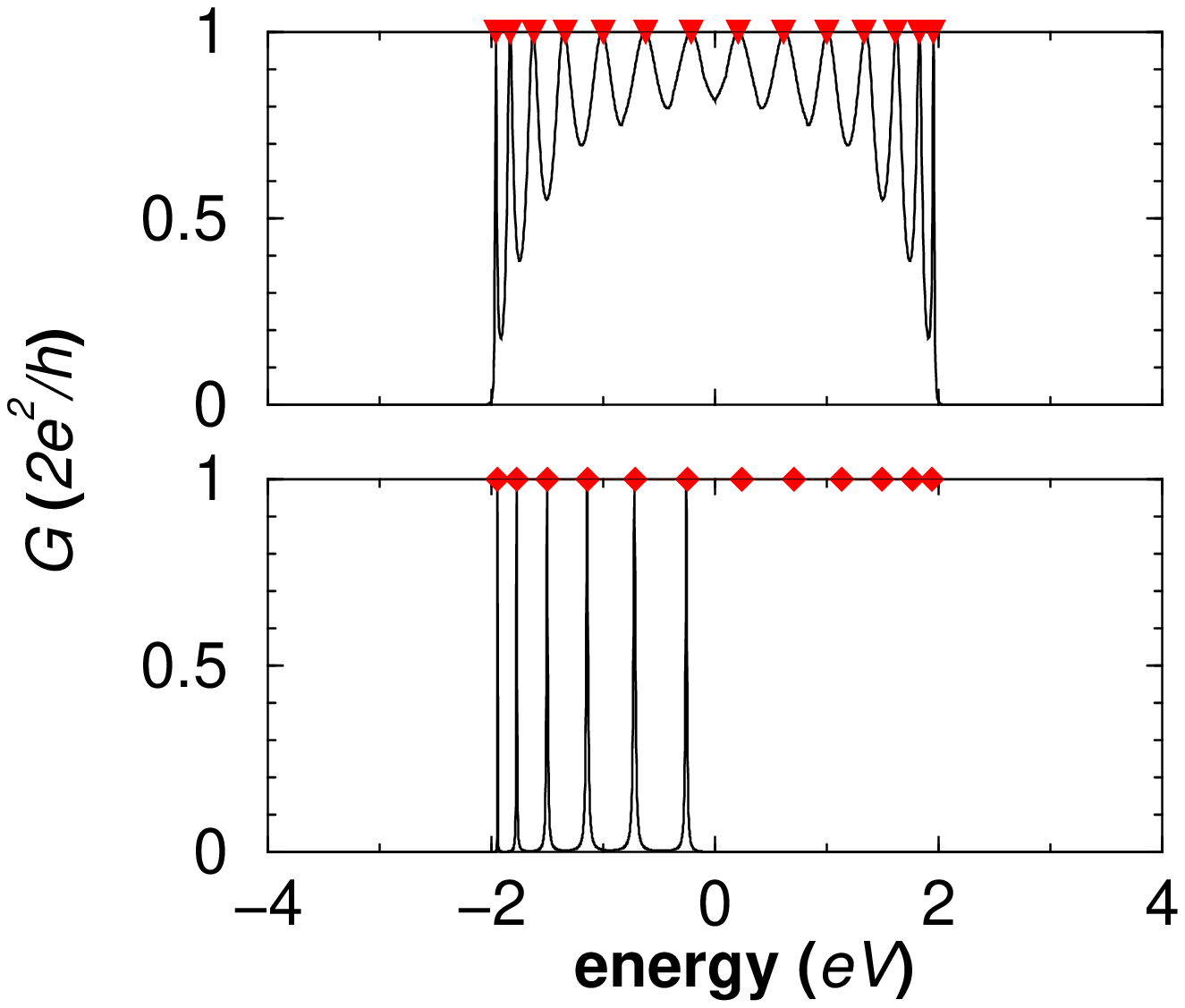, width=.49\linewidth}}
\subfigure{\epsfig{file=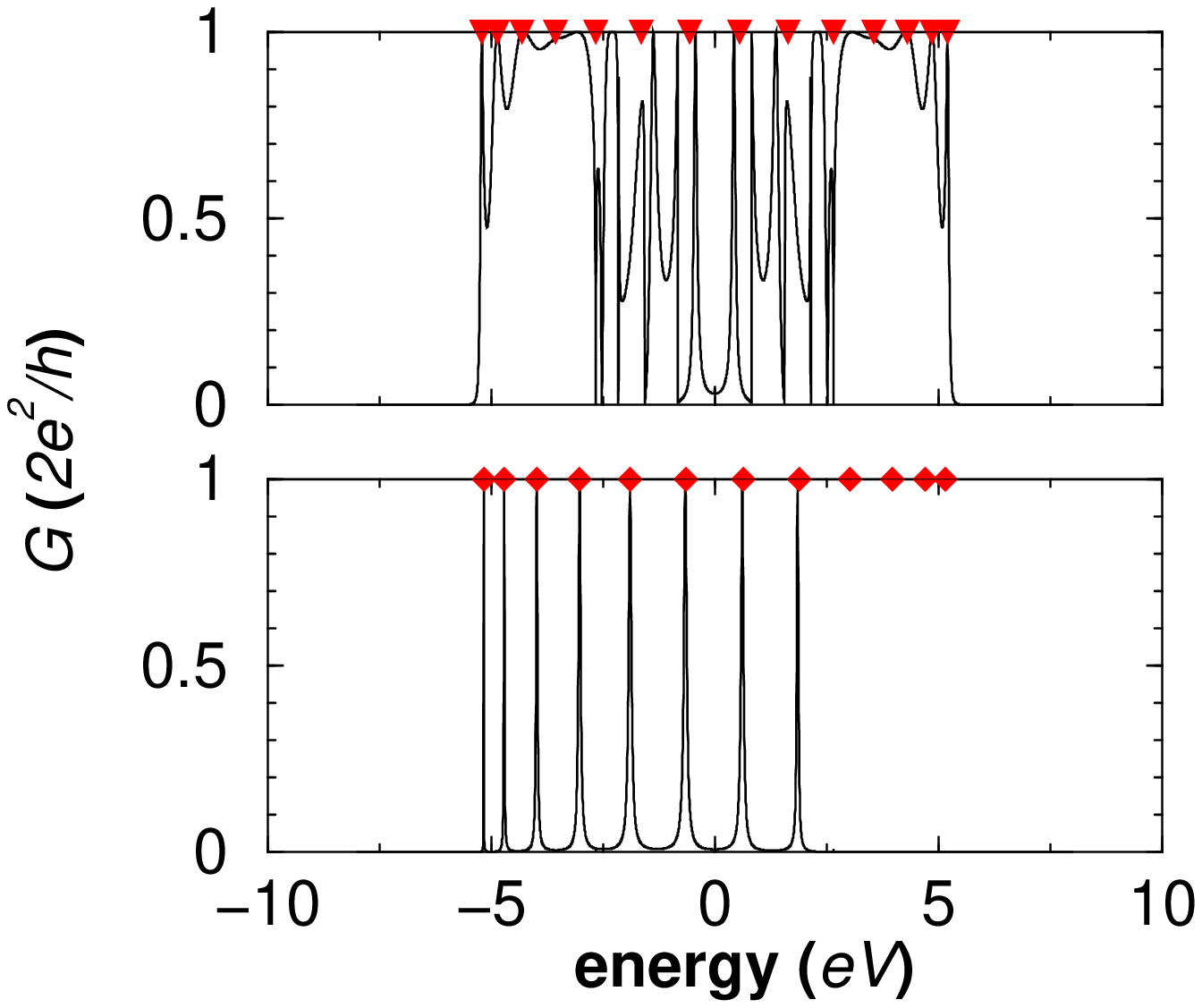, width=.49\linewidth}}
\end{center}
\vspace*{-5mm} 

\caption{\label{fig1}
	Typical conductance spectra for an $N\!=\!14$ molecular bridge between
	square-lattice tube (left) and (10,10) carbon nanotube 
	electrodes (right). An effective multiple contact ($M_c\!=\!20$, bottom)
	acts as a filter selecting a single transport channel in contrast
	to a single contact (top). 
	Diamonds and triangles indicate eigenenergies of an $N\!=\!12$ and $N\!=\!14$
        iso\-la\-ted chain. $\varepsilon^{\rm wire}\!\!=\!0$ and
	$\gamma^{\rm L,R}\!=\!\gamma^{\rm wire}\!=\!\Gamma$.\ The carbon nanotube,
	square-lattice tube, and molecular wire bands (centered at
	zero) have widths 16eV, 8eV, and
	$4\gamma^{\rm wire}$, respectively.
	}
\end{figure}
Gross properties of the conductance spectrum of the system can be understood
by looking at the two extreme cases of a single interfacial contact,
$M_{\rm c}=1$~(single contact-SC), and multiple  contacts, $M_{\rm c}=M$
~(MC). In the SC-scenario 
all open channels contribute to the transmission, i.e.\
$S_{\nu\nu^\prime}(E)$ is non-zero for any $\nu,\nu^\prime$.
For the case of the molecular wire bridging two square-lattice tubes,
depicted in the upper left panel of Fig.~\ref{fig1}, 
the conductance bears some of the properties
of current flowing through a one-dimensional constriction~\cite{FG99}.
In particular, the conductance shows resonances of quantum unit~($2 e^2/h$)
height at eigenenergies of the isolated molecular chain
(indicated as triangles).
They arise because of back reflections at the molecular interface. However,
for carbon tubules leads~(upper right panel of Fig.~\ref{fig1}) we observe
additional structure in the conductance spectra.
Preliminary results suggest that distinctive features such as
antiresonances are signatures of
van Hove singularities in the
carbon tubules band structure~\cite{preliminary}.

In contrast, the MC-configuration acts as a channel filter resulting in a profoundly different
behavior:  The conductance vanishes for part of the spectrum 
as shown in the lower panels of Fig.~\ref{fig1}.
The complicated 
conductance spectrum for a SC carbon tubule-molecule configuration turns into 
a regular sequence of peaks at eigenenergies of the isolated molecular wire
(marked as diamonds).
Further analysis of the $S$-matrix elements revealed
that only
wavefunctions of the tubes without modulation along the
cross-section circumference allow for transport, thereby, yielding zero conductance
when such channels are not available. The filtering is a consequence 
of a sum rule that determines the transmission of each open channel which may be viewed as the
overlap $\langle n_{\rm wire}| H| n_{\rm L,R} \rangle$
(see Eq.~(\ref{meq1}), due to the nature of coupling only the transverse profile is important).
The overlap is also related to the spectral density~\cite{YR98}. 
For the square lattice model tubes e.g. only the channel with
zero `transverse' momentum gives a non-zero sum. 
From symmetry considerations it is clear
that the channel selection is a generic feature of cylindrical electrodes.
Moreover, we found that channel filtering approximately prevails also 
for non-cylindrical, mesoscopic electrodes with lateral confinement.
More generally, multiple
contacts allow for control of low-dimensional transport via channel selection.
\vspace*{-5mm} 

\begin{figure}
\begin{center}
\subfigure{\epsfig{file=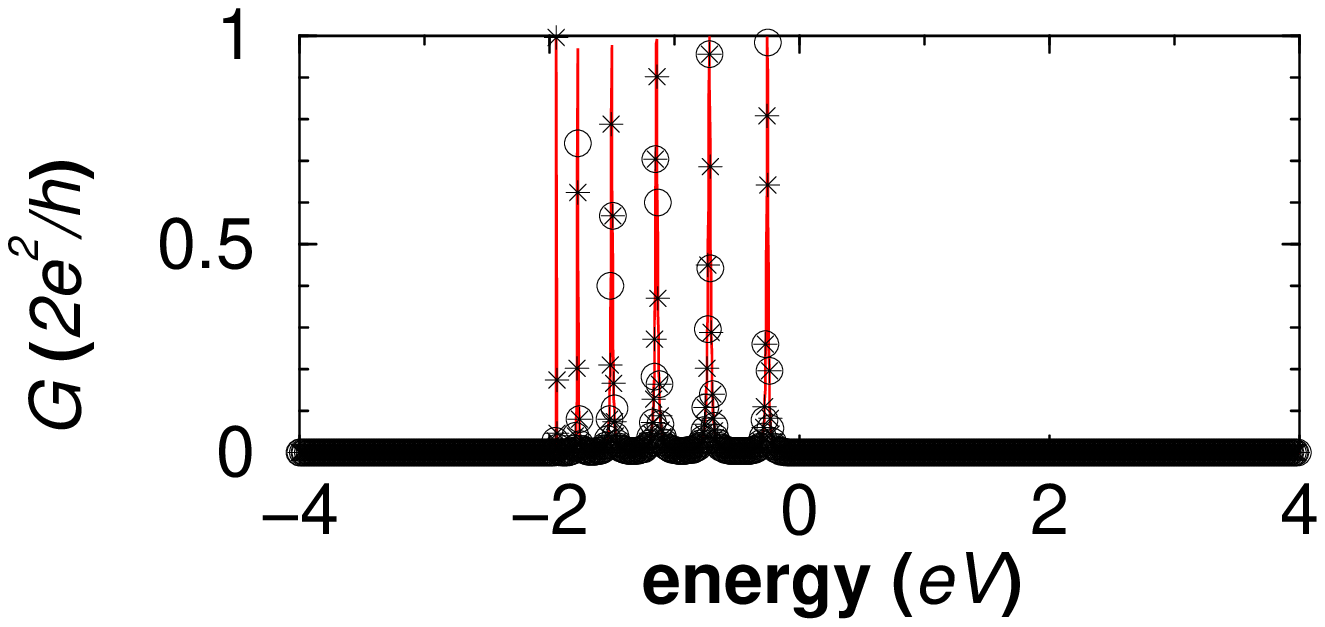, width=.49\linewidth}}
\subfigure{\epsfig{file=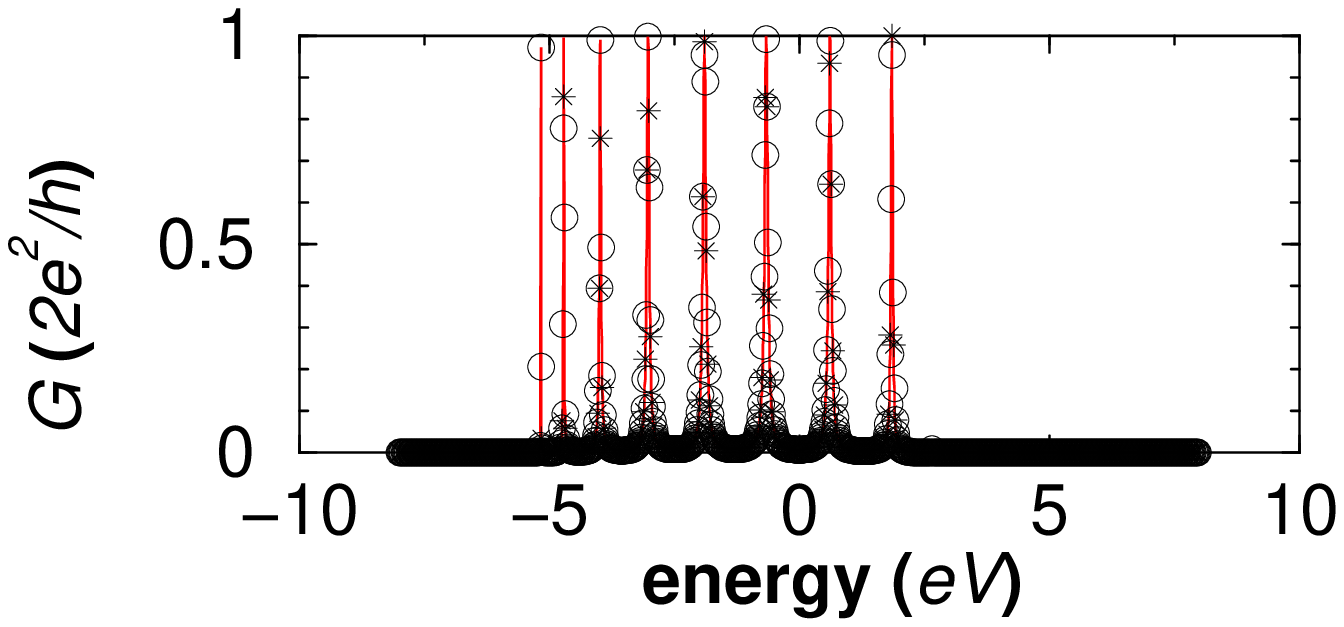, width=.49\linewidth}}
\end{center}
\vspace*{-5mm} 

\caption{\label{fig2}
	Superimposed conductance curves with $\varepsilon^{\rm wire}=0$,
	$\gamma^{\rm L,R}=\gamma^{\rm wire}$, and $\Gamma^2 M =$ const
	showing the validity of the sum rule (see text) for the multiple
	contact configuration
	for square lattice tube (left) and carbon nanotube (right) electrodes.
	Symbols and line indicate different $\Gamma$.}
\end{figure}

An additional particular feature of the MC configuration is that
the conductance conforms to a scaling law.
Two MC hybrid--structures differing in tube diameters $D$ and contact
strengths $\Gamma$, but conserving the product $\Gamma^2 \cdot D$, exhibit the
same conductance profile~(Fig.~\ref{fig2}). This is a mere contact effect
related to the symmetry of the contributing channel wavefunction and should
hold for any effective coupling with the form considered here. For 
single-channel transport  it follows that the conductance is 
proportional to $|\langle n_{\rm wire}| H| n_{\rm L,R} \rangle|^2$. Then one
readily arrives at the exact form of the scaling law taking into account that
the transverse profile of the contributing channel wavefunction has no nodes and is
normalized.
\vspace*{-5mm} 
\begin{figure}[t]
\begin{center}
\subfigure{\epsfig{file=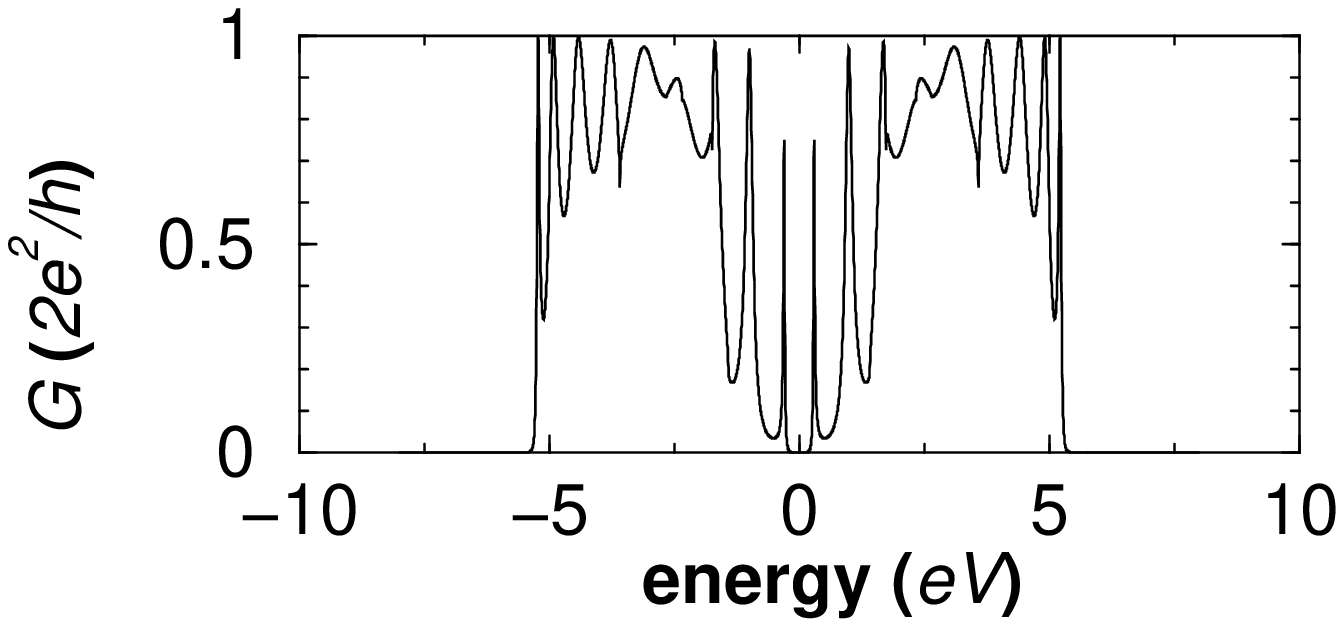, width=.49\linewidth}}
\subfigure{\epsfig{file=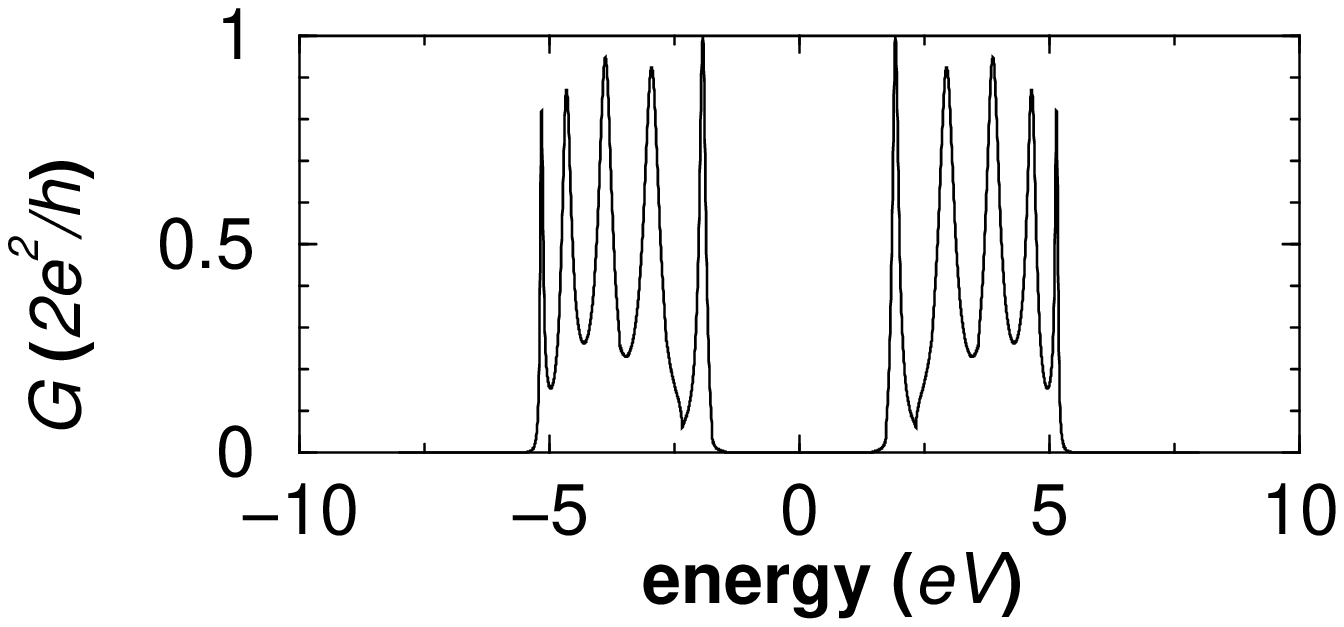, width=.49\linewidth}}
\end{center}
\vspace*{-5mm} 

\caption{\label{fig3}
	Conductance spectrum for (9,0) carbon nanotube electrodes with $M_c=1$ (left)
	and $M_c=3$ (right). $\varepsilon^{\rm wire}=0$ and
	$\gamma^{\rm L,R}=\gamma^{\rm wire}=\Gamma$.
	}
\end{figure}
The `intermediate contact' case, i.e.~$1\!<\!M_c\!<\!M$, exhibits much
richer behavior but can be understood with the above arguments. For
completeness we discuss a specific example which once more
illustrates the importance of the interfacial coupling
for molecular systems bridging nanoelectrodes and further supports channel
selection. In Fig.~\ref{fig3} typical
conductance spectra for zigzag carbon nanotube electrodes are shown. To interpret these results
we note that transport usually takes place at $E\sim1$eV around the Fermi
energy $E_F=0$. For this part of the spectrum we notice a complete suppression of conductance
for $M_c=3$, owing to contact `dimensionality' \cite{f2}.
The origin of this effect derives from metallic zigzag nanotubes supporting
two degenerate transport channels in this energy region with wavefunction symmetries such
that the wire/tube overlap gives a zero contribution for $M_c=3 \cdot n$ as
depicted and non-zero otherwise \cite{f3}.
\vspace*{-5mm} 

\begin{figure}[t]
\begin{center}
\subfigure{\epsfig{file=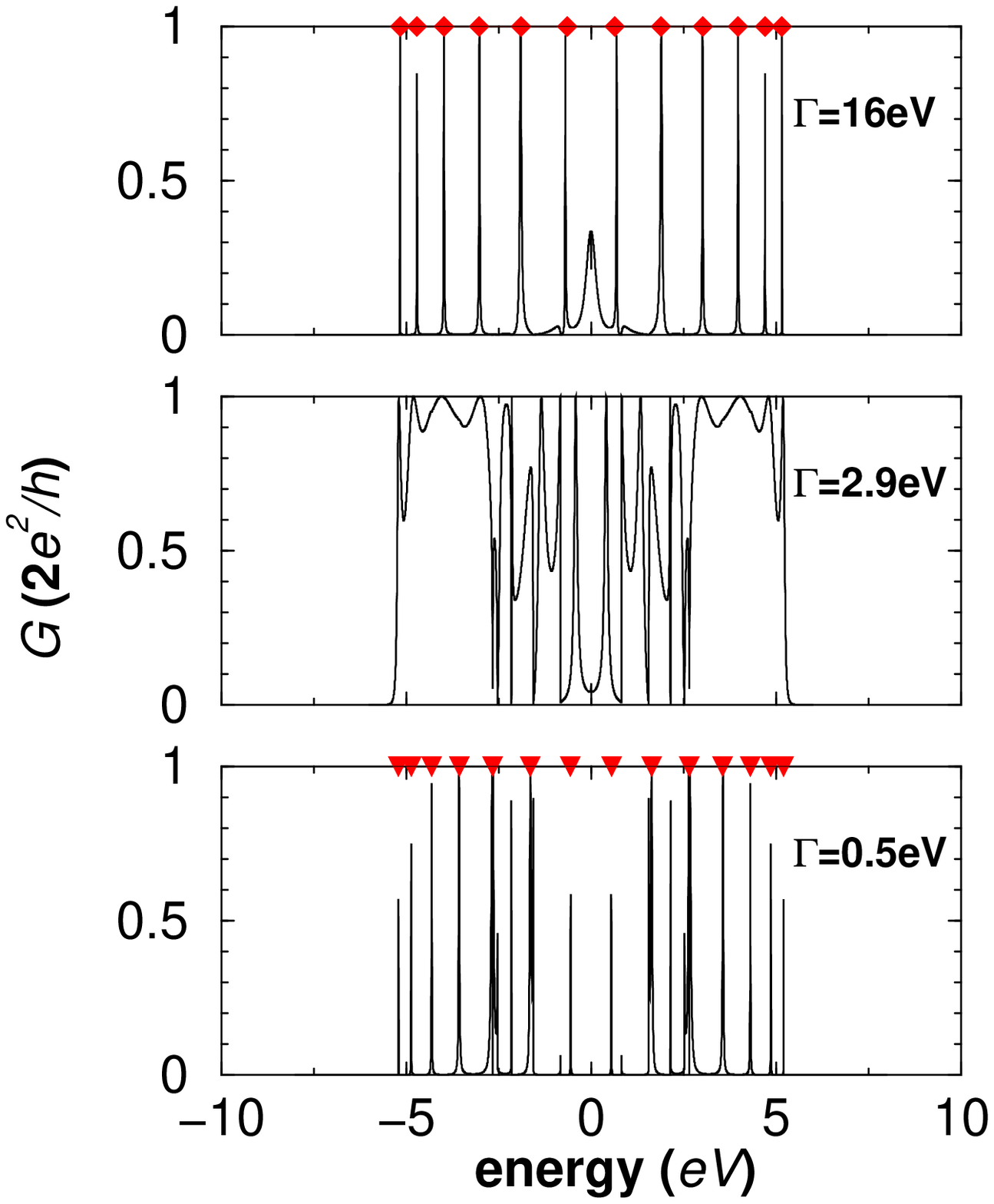, width=.49\linewidth}}
\subfigure{\epsfig{file=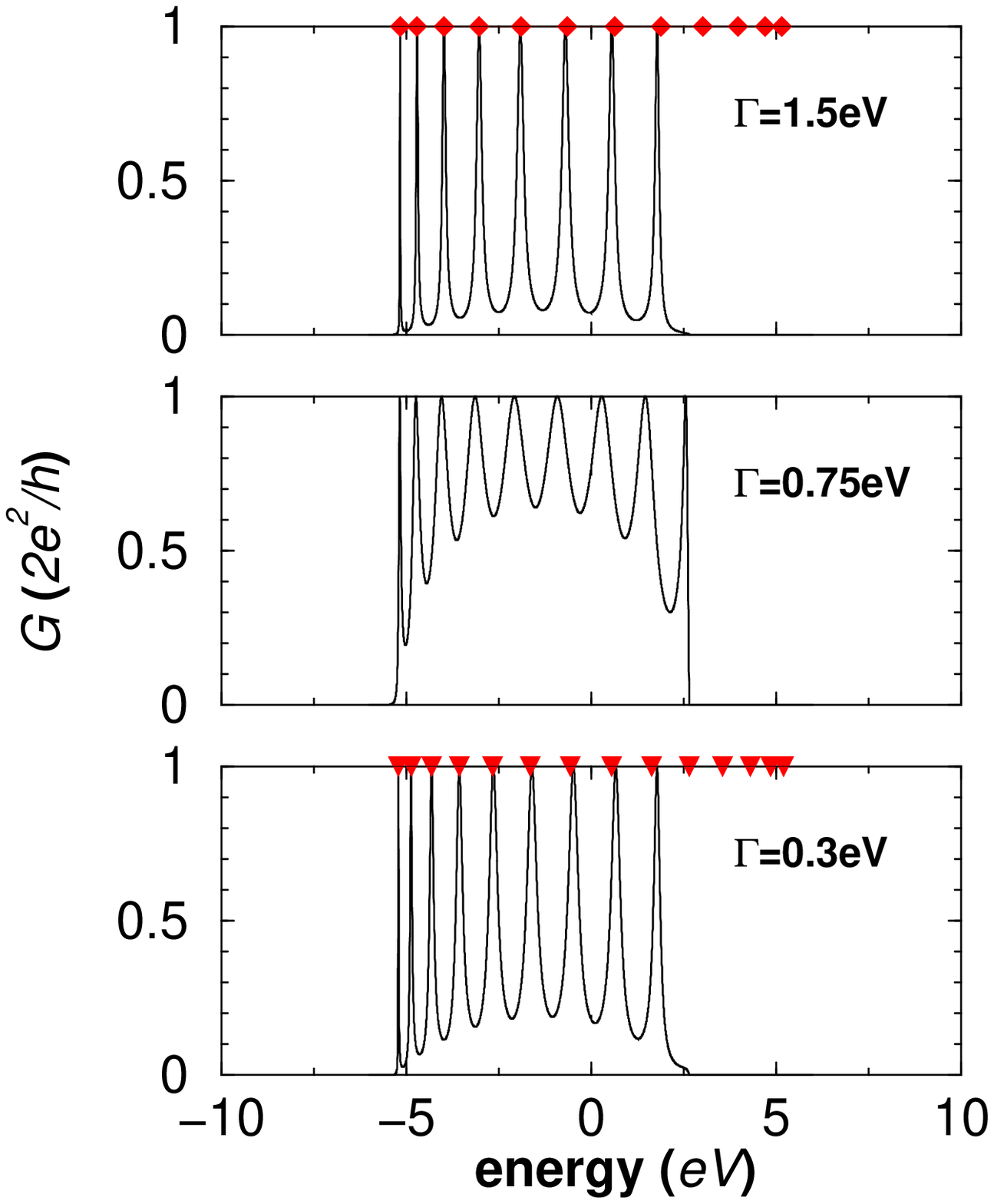, width=.49\linewidth}}
\end{center}
\vspace*{-5mm} 

\caption{\label{fig4}
	Chemical binding effects: the effective length of the molecular bridge~($N=14$)
	depends on the contact strength. Conductance resonances are
	followed by a narrow-to-broad-to-narrow behavior
	as $\Gamma$ decreases from top to bottom for both
                  single~(left) and multiple contact~($M_c=20$, right). Diamonds and
	triangles indicate eigenenergies of an $N=12$ and $N=14$ isolated chain,
	respectively. $\varepsilon^{\rm wire}=0$, $\gamma^{\rm L,R}=\gamma^{\rm wire}$,
	and $\Gamma$ is given as an inset.
	}
\end{figure}
Finally, we examine the dependence of the conductance on the coupling strength $\Gamma$.
It is expected that as the coupling is reduced,
resonances are resolved with decreasing width-to-height ratio. This is indeed
visible when comparing the middle and lower panels of Fig.~\ref{fig4} for both single~(left) and
multiple contact~(right). However, the upper panels of Fig.~\ref{fig4} reveal a counterintuitive
behavior. By decreasing $\Gamma$ from top to bottom in Fig.~\ref{fig4} we  observe 
an intermediate broadening of the resonance widths, accompanied by a clear-cut change
in the position and number of resonances picked up. There is a crossover from
$N-2 \rightarrow N$ resonances, if all the wire resonant states (marked as diamonds and triangles
in Fig.~\ref{fig4}) fall into the tube energy band.
The position of resonances is suggestive
for a possible explanation of this effect. It points to considering the wire end-atoms as part of the
leads due to strong binding with the transition driven by the interaction strength.
Similar binding effects, which determine the effective length (or the nature) of the molecular
bridge, have been discussed in Ref.~\cite{EK99a} and are quite sensitive to
multiple contacts as demonstrated.

In additional simulations \cite{preliminary},
we have considered a dimerized chain as a
model for a molecular wire as suggested in~\cite{JV96} and more complex artificial molecular
compounds, which nevertheless do not show distinct behavior from what we have discussed.

We have not discussed electron transport for finite temperatures and bias
voltages, since our objective has been to point out
generic geometry-induced features linked to the contact resistance of molecular wires.
In that case the calculation of $I$-$V$ characteristics requires
generalized methods with the Landauer conductance delivering the initial insight.
We further point out that in the scattering approach used it is implicitly assumed
that no inelastic scattering or other phase-breaking mechanisms occur. For instance,
coupling to molecular vibrational modes is neglected and, therefore, studies relying on the
independent-electron picture may be only indicative of the underlying
conduction mechanisms. We have also assumed that the molecular system is strongly
coupled to the electrodes so that charging effects are not important.

In summary, we have studied contact effects on the
conductance of a hybrid built by a molecular wire coupled to low-dimensional leads.
We have pointed out immediate consequences of the contact geometry and dimensionality
on electron transport across such systems.
These include channel selection, conductance spectra tuning, and the existence of
a scaling law $G=G(\Gamma^2\! \cdot\! M)$. We also demonstrated that a square lattice tube
exhibits most of the above features and can be used as a tractable analytic model.
Replacing such a model for electrodes by natural and realistic candidates of molecular electronic circuits,
carbon nanotubes, adds to the conductance profile richer structure. The latter needs further
investigation 
to relate it to the tubules band structure and details depending on the tubule
chirality. 
We hope that our observations may drive systematic chemical synthesis based on
geometrical factors and stimulate corresponding experimental analysis. 
Especially for multi-wall nanotube leads, an axial magnetic field of reasonable 
magnitude can be applied which modulates the symmetry of the
tube states and, hence, can act as an external tuning parameter.

We acknowledge fruitful discussions with A. Bachtold, V.I. Fal'ko, and P. Fulde. 
G.C. was partially supported by the EU under contract TMR-ERBFMRXCT980180.

\ecols

\begin{thebibliography}{10}
\bibitem{f2}
The position of the contacts does not influence conductance for zigzag carbon 
tubule leads, while for armchair carbon tubules there is no distinction only if $M_ c$ is odd.
\bibitem{f3}
For armchair
tubules,
conductance is always finite but for $M_c$ even,
only one channel instead of two is contributing.
\end{thebibliography}
\end{document}